\RequirePackage{fix-cm}
\documentclass[smallextended]{svjour3}       
\smartqed  
\usepackage{amsmath,bbold}
\usepackage{amsfonts,tikz}

\usepackage{graphicx}
\begin{document}

\title{A quenched 2-flavour Einstein-Maxwell-dilaton gauge-gravity model\thanks{The work of S.M.~is supported by the Department of Science and Technology,
Government of India under the Grant Agreement number IFA17-PH207 (INSPIRE Faculty Award).
}
}


\author{D.~Dudal        \and
        A.~Hajilou \and S.~Mahapatra 
}


\institute{D.~Dudal \at
              KU Leuven Campus Kortrijk--Kulak, Department of Physics, Etienne Sabbelaan 53 bus 7657, 8500 Kortrijk, Belgium \\
              Ghent University, Department of Physics and Astronomy, Krijgslaan 281-S9, 9000 Gent, Belgium
              \email{david.dudal@kuleuven.be}           
           \and
           A.~Hajilou \at
           School of Particles and Accelerators, Institute for Research in Fundamental Sciences (IPM), Tehran 19395-5531, Iran
              \email{hajilou@ipm.ir}
           \and
           S.~Mahapatra \at
              Department of Physics and Astronomy, National Institute of Technology Rourkela, Rourkela-769008, India
              \email{mahapatrasub@nitrkl.ac.in}
}

\date{Received: date / Accepted: date}

\maketitle

\begin{abstract}
We extend earlier work by introducing an Einstein--Maxwell--dilaton (EMD) action with two quark flavours. We solve the corresponding equations of motion in the quenched approximation (probe quark flavours) via the potential reconstruction method in presence of a background magnetic field in search for a self-consistent dual magnetic AdS/QCD model. As an application we discuss the deconfinement transition temperature confirming inverse magnetic catalysis, whilst for moderate values of the magnetic field also the entropy density compares relatively well with corresponding lattice data in the vicinity of the transition.



\end{abstract}

\section{Introduction}
\label{intro}
Besides the high temperatures reached in heavy ion collisions, a strong magnetic field is generated in peripheral collisions \cite{Skokov:2009qp,Tuchin}.  Although the field due to the spectator particles decays very quickly, the resulting magnetic field can be considered roughly constant due to the medium response, being usually modelled via $\vec{B}=B\vec{e}_3$ with $B$ constant. The maximum strength of this field is estimated to be around $10^{16}$ Tesla.  Since quarks carry, next to their color charge, also electric charge, their dynamics is thus strongly susceptible to a magnetic field. In particular, it must influence the QCD phase diagram, figuring as another external parameter.
This has induced an increased research activity in the area of magnetized QCD, let us refer to the recent review works \cite{tomes}.

Despite the fact that the temperature and magnetic field are immensely large, their values in units of the fundamental QCD scale $\Lambda_{\textrm{QCD}}$ are still such that the dynamics is strongly coupled, making its study challenging and a rich field of ongoing research.

One fruitful way to study strongly coupled magnetic QCD is via lattice QCD, since adding a magnetic field to the game does not lead to a sign problem, so Monte Carlo sampling remains possible, see e.g.~\cite{Bali:2011qj,Bali:2012zg}. Next to this, one can also resort to effective low energy QCD models based on its global symmetries and relevant order parameters, see e.g.~\cite{Mizher:2010zb,Ferreira:2014kpa,Fukushima:2012kc}, or using the gauge-gravity paradigm, considering strongly coupled QCD to be dual to an appropriate semi-classical gravity model, see f.i.~\cite{Callebaut:2013ria,Rougemont:2015oea,Mamo:2015dea,Dudal:2015wfn,Giataganas:2017koz,Gursoy:2017wzz,Rodrigues:2018pep,Bohra:2019ebj,Bohra:2020qom,Ballon-Bayona:2020xtf} for a few examples.

\section{Model setup}
The gravity action we propose is given by a five-dimensional EMD gravity, see also \cite{Bohra:2019ebj},
\begin{eqnarray}
S_{EM} &=&  -\frac{1}{16 \pi G_5} \int \mathrm{d^5}x \sqrt{-g}  \ \left[R - \frac{f(\phi)}{4}\textrm{Tr}[F_{(L)MN}F_{(L)}^{MN}+F_{(R) MN}F_{(R)}^{MN}]\right.\nonumber\\&&\left. -\frac{1}{2}\partial_{M}\phi \partial^{M}\phi -V(\phi)\right],
\label{actionEMD}
\end{eqnarray}
where $G_5$ is the Newton constant in five dimensions, $R$ is the Ricci scalar, $\phi$ is the dilaton field, $F_{(L)MN}$ and $F_{(R)MN}$ are the field strength tensors for the two $U(2)$ flavour gauge fields, $f(\phi)$ is the gauge kinetic function that acts as coupling between the gauge fields and the dilaton field and $V(\phi)$ is the dilaton potential.

The left and right gauge fields can be combined to describe the two-flavour vector and axial currents ($SU(2)\times SU(2)$-sector), next to the baryon and anomalous axial currents ($U(1)\times U(1)$-sector). For the current purposes, we will only be interested to source the theory with a background magnetic field coupling to the (probe) charge $+2/3$ up and charge $-1/3$ down quark. Concretely, we can slightly gauge \cite{Sakai:2004cn,Callebaut:2013ria} the $U(1)\times U(1)$ symmetry, leading to a vector magnetic background $V_M = A_{(L)M}+A_{(R)M}$. Taking
\begin{equation}\label{vm}
  V_M = \left(x_1 B \delta_{M,2} \frac{\mathbb{1}}{6}+x_1 B \delta_{M,2} \frac{\sigma_3}{2}\right) 
\end{equation}
yields
\begin{equation}\label{vm2}
  F_{12} =  B\left(
               \begin{array}{cc}
                 \frac{2}{3} & 0 \\
                 0 & -\frac{1}{3} \\
               \end{array}
             \right)\equiv  B \mathbb{Q}
\end{equation}
where $x_1$ is the first spatial boundary coordinate (see eq.~(\ref{ansatze}) for more details).  

In foregoing work like \cite{Bohra:2019ebj,Bohra:2020qom,Arefeva:2018cli}, adding the magnetic field was achieved via a simplified $U(1)\times U(1)$ setup, with two (independent) scale factors $f_1(\phi)$ and $f_2(\phi)$. One of them followed from the equations of motion, another one could be chosen freely to capture the QCD vector meson mass spectrum. The here proposed (closer to QCD) generalization with a $U(2)\times U(2)$ automatically leads to less freedom, as the now single coupling $f(\phi)$ will be fixed by the equations of motion. Elsewhere, we will investigate to what extent the vector meson spectrum is still reasonably described by this dynamically fixed $f(\phi)$.

To obtain the on-shell solutions, the following Ans\"{a}tze have been considered for the metric field $g_{MN}$ and dilaton field $\phi$, next to \eqref{vm2},
\begin{eqnarray}
 ds^2=\frac{L^2 S(z)}{z^2}\biggl[-g(z)dt^2 + \frac{dz^2}{g(z)} + e^{B^2 z^2} \biggl( dx_{1}^2 + dx_{2}^2 \biggr) +dx_3^2\biggr]\,, \quad \phi=\phi(z),
\label{ansatze}
\end{eqnarray}
where $L$ is the AdS length scale, $S(z)$ a to be chosen scale factor, and $g(z)$ is the blackening function. Thermal AdS will correspond to $g(z)\equiv 1$. Here, $z$ is the radial coordinate with $z = 0$ at the AdS boundary. This coordinate $z$ runs from the boundary to the horizon at $z=z_h$ for the black hole case or to $z=\infty$ for the thermal AdS case. The boundary conditions, consistent with asymptotically AdS-behaviour, are given by $g(0)=1$, $g(z_h)=0$, $S(0)=1$, $\phi(0)=0$. Notice that we use here a five-dimensional bulk magnetic field $B$ (mass dimension one). By eventually rescaling $B\to B L$, we can work with a mass dimension two magnetic field, as appropriate for four dimensions.

The Euler-Lagrange equations then lead to, upon setting $\mathcal{B}^2=B^2 \textrm{Tr}[\mathbb{Q}^2]$,
\begin{eqnarray}
g''(z) +g'(z) \left(2 B^2
   z+\frac{3 S'(z)}{2 S(z)}-\frac{3}{z}\right)  = 0,
\label{EOM11}
\end{eqnarray}
\begin{eqnarray}
\frac{2 \mathcal{B}^2 z e^{-2 B^2 z^2} f(z)}{L^2 S(z)}+2 B^2 g'(z)+g(z)
   \left(4 B^4 z+\frac{3 B^2 S'(z)}{S(z)}-\frac{4 B^2}{z}\right) = 0,
\label{EOM22}
\end{eqnarray}
\begin{eqnarray}
S''(z)-\frac{3 S'(z)^2}{2 S(z)}+\frac{2 S'(z)}{z} + S(z) \left(\frac{4 B^4 z^2}{3}+\frac{4 B^2}{3}+\frac{1}{3} \phi
   '(z)^2\right) = 0,
\label{EOM33}
\end{eqnarray}
and, finally,
\begin{eqnarray}
&&\frac{g''(z)}{3g(z)} +\frac{S''(z)}{S(z)} +\frac{S'(z)}{S(z)} \left(\frac{7 B^2 z}{2 }+\frac{3 g'(z)}{2 g(z)
   }-\frac{6}{z }\right) + g'(z) \left(\frac{5 B^2 z}{3
   g(z)}-\frac{3}{z g(z)}\right)  \nonumber \\
     &&+ 2 B^4 z^2+\frac{\mathcal{B}^2 z^2 e^{-2 B^2 z^2} f(z)}{3 L^2 g(z)
   S(z)}-6 B^2+\frac{2 L^2 S(z) V(z)}{3 z^2
   g(z)}+\frac{S'(z)^2}{2 S(z)^2}+\frac{8}{z^2} = 0.
\label{EOM44}
\end{eqnarray}
There is also the dilaton equation, which is however not independent w.r.t.~the above equations, hence we do not record it here. Moreover, the gauge field equation of motion is trivially satisfied for our constant magnetic background.

Utilizing the above Ans\"{a}tze eq.~(\ref{ansatze}), imposing suitable boundary conditions and following the procedure outlined in \cite{Bohra:2019ebj}, complete solutions can be expressed in terms of a single arbitrary function $S(z)$.

Solving the above equations \eqref{EOM11}-\eqref{EOM44} in the order they stand, with $S(z)=e^{2A(z)}$, consecutively leads to
\begin{eqnarray}
g(z) &=& 1 -\frac{1}{\int_0^{z_h} \, d\xi \ \xi^3 e^{-B^2 \xi^2-3A(\xi)} } \int_0^z \, d\xi \ \xi^3 e^{-B^2 \xi^2 -3A(\xi) },
  \label{gsol}
\end{eqnarray}
\begin{eqnarray}
f(z) = - \frac{ e^{2 B^2 z^2 + 2 A(z)} L^2}{z\textrm{Tr}[{\mathbb Q}^2]}\biggl[ g(z) \left(2 B^2 z+
   3A'(z)-\frac{2}{z}\right)+ g'(z) \biggr],
\label{f2sol}
\end{eqnarray}
and
\begin{eqnarray}
\phi(z) &=& K\\&&\hspace{-1.5cm}+ \int \, dz  \frac{\sqrt{2} \sqrt{3 z A'(z)^2-6 A'(z)-3 z A''(z)-2 B^4 z^3-2 B^2 z}}{\sqrt{z}} \nonumber
\label{phisol}
\end{eqnarray}
with $K$ uniquely determined from the boundary condition $\phi |_{z=0}= 0$. At last, the potential can be extracted from
\begin{eqnarray}
V(z) & = & \frac{3 z^2 g(z)}{2  e^{2 A(z)}L^2} \left(-\frac{\mathcal{B}^2 z^2 f(z) e^{-2 B^2 z^2}}{3 L^2 e^{2 A(z)} g(z)}-\frac{g'(z)}{g(z)} \left(3 A'(z)+\frac{5 B^2 z}{3}-\frac{3}{z
  }\right) \right) \nonumber\\
 & & +  \frac{3 z^2 g(z)}{2  e^{2 A(z)}L^2}  \left(-7 B^2 z A'(z)+\frac{12 A'(z)}{z}-2 B^4 z^2+6
   B^2-\frac{8}{z^2}\right)  \nonumber\\
  & &   + \frac{3 z^2 g(z)}{2  e^{2 A(z)}L^2} \left( -2 A''(z) -\frac{g''(z)}{3 g(z)}  -6 A'(z)^2           \right)   .
\label{Vsol}
\end{eqnarray}
The black hole temperature and entropy density are determined as
\begin{eqnarray}\
T =   \frac{z_h^3 \ e^{-3 A\left(z_h\right)-B^2 z_h^2}}{4 \pi\int_0^{z_h} \, d\xi \ \xi^3 e^{-B^2 \xi^2-3A(\xi)}}, \quad s = \frac{e^{B^2 z_{h}^{2}+3A(z_h)} L^3}{4G_5 z_{h}^3}.
\label{phicase11}
\end{eqnarray}
In \cite{Bohra:2019ebj,Bohra:2020qom}, we discussed the self-consistency of the utilized potential reconstruction method \cite{Alanen:2009xs} in terms of the Gubser stability criterion \cite{Gubser:2000nd}. Furthermore, it can be explicitly verified that the (on-shell) potential $V(z)$ is virtually independent of the temperature and/or magnetic field.

The above construction works for every choice of form factor $A(z)$. We will hence choose it ``wisely'', that is, sufficiently simple for computations but also capable of describing QCD-features qualitatively well. We will adapt the same choice, $A(z)= -az^2 - d B^2 z^5$, as in \cite{Bohra:2020qom}, with $a=0.15~\textrm{GeV}^2$ and $d=0.013~\textrm{GeV}^3$. The $z^2$-term leads to a reasonable deconfinement temperature in the $B=0$ case\footnote{Concretely, we have $T_\textrm{crit}=0.268~\text{GeV}$, in the ballpark of \cite{Lucini:2003zr}.},  while the extra $z^5$-term is necessary to avoid an unphysical breaking of the interquark string in the magnetized case, remembering that we are in the quenched case, so there are no dynamical quarks to induce such a string breaking. As the dilaton solution ought to be real, our model is restricted to values of $BL< 1.02~\textrm{GeV}^2$.

Analogously as analyzed in \cite{Bohra:2019ebj}, one can show that there are actually two black hole solutions per choice of temperature, however only one of these is thermodynamically stable.

\section{Deconfinement phase transition}
\begin{figure}
  \includegraphics[width=8cm]{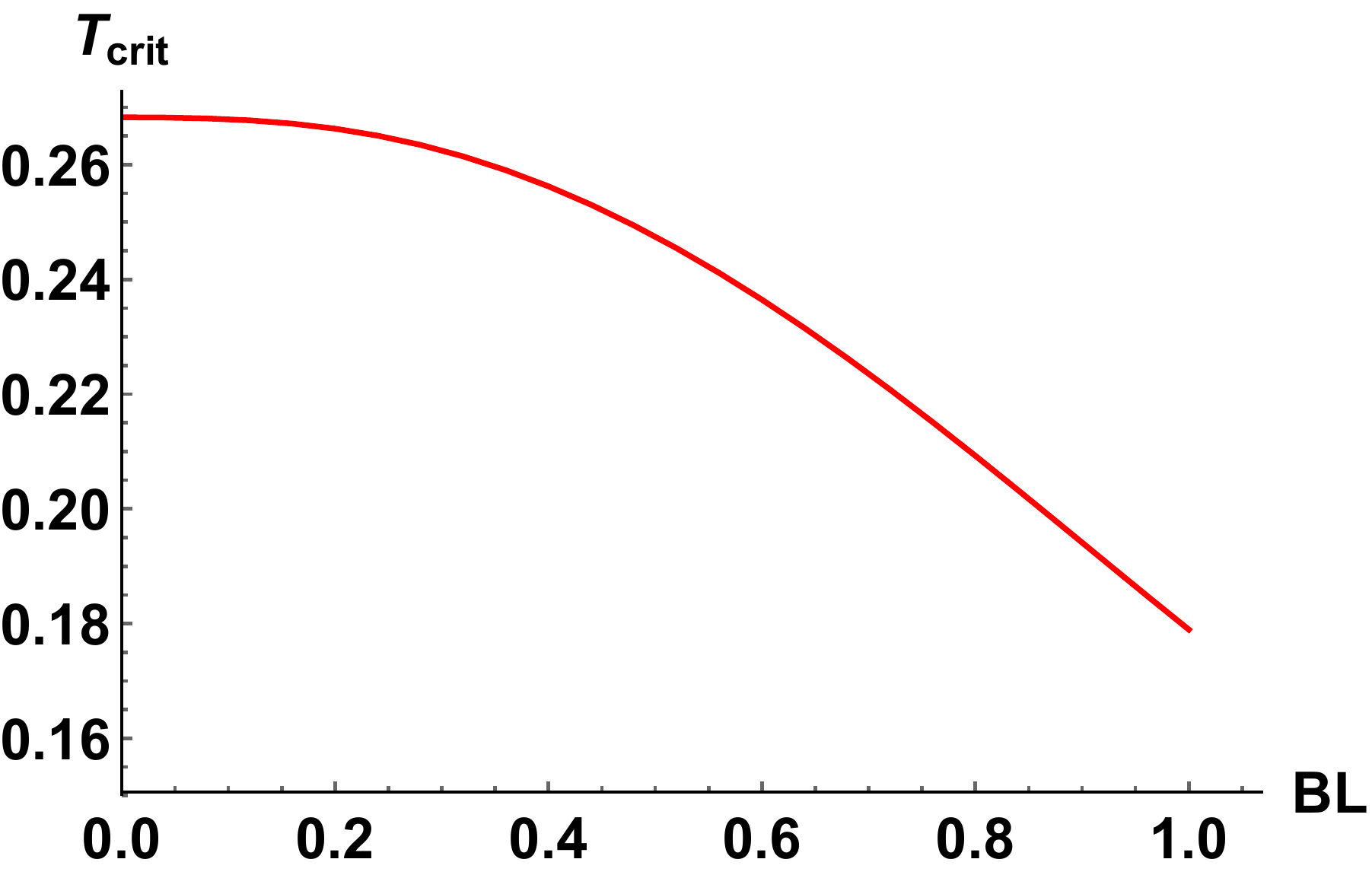}
\caption{Deconfinement transition temperature in terms of magnetic field. In units GeV.}
\label{deconf}       
\end{figure}

Let us first have a look at the deconfinement temperature that we can extract from its holographic dual, the Hawking--Page temperature $T_{\textrm{crit}}$ corresponding to the thermodynamical phase transition from the black hole to the thermal (non-black hole) case. The latter temperature follows from comparing the free energies of both cases computed from the action \eqref{actionEMD}. Doing so leads to Fig.~\ref{deconf}, indeed indicative of inverse magnetic catalysis.

\section{Entropy density}
Another interesting non-perturbative quantity is the quark-gluon plasma entropy density $s$, which we can compare with \cite[Fig.~10]{Bali:2014kia}. In order to do so, let us first re-express the Newton constant $G_5$ in terms of $L$, working at $B=0$. This can be done by matching the free energy in the high temperature limit $T\to\infty$ (or $z_h\to0$) with the standard ``free'' Yang-Mills result (free gluon gas obeying a Stefan-Boltzmann law). Using \eqref{phicase11} in combination with $S_{\textrm{free}} = \frac{4(N^2-1)}{45} \pi^2 T^3$ leads to $G_5=\frac{45\pi L^3}{16(N^2-1)}$. Notice that for $z_h\to0$, we have $T\to\frac{1}{\pi z_h}$. This $G_5$-value coincides with that of \cite{Megias:2010ku}, since their UV limit matches asymptotically with our $B=0$ model when we consider $z_h\to0$.

\begin{figure}
  \includegraphics[width=8cm]{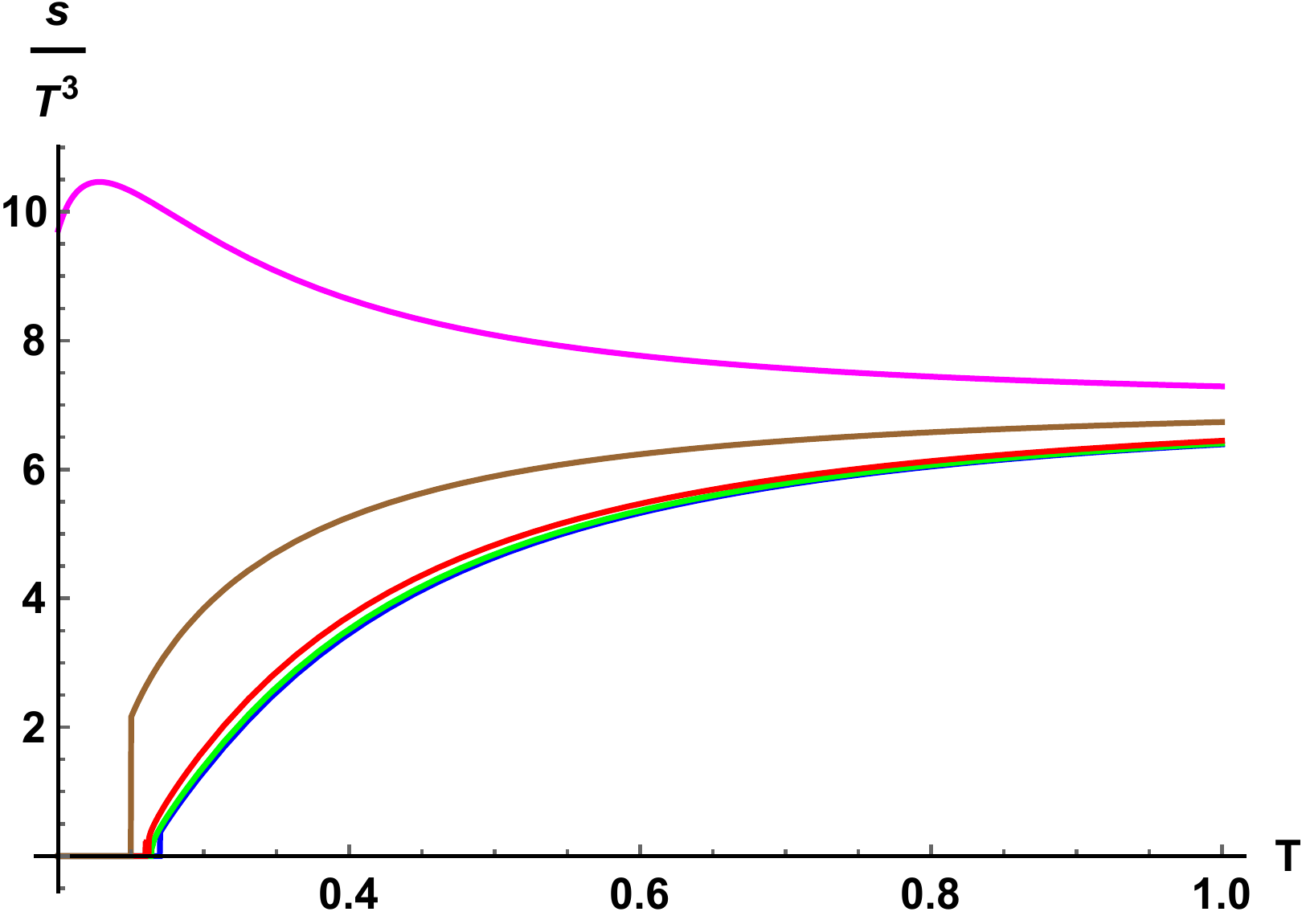}
\caption{ Entropy density normalized by $T^3$ for $BL=0$~(blue), $BL=0.1$~(green), $BL=0.2$~(red), $BL=0.5$~(brown), $BL=0.8$~(magenta). In units GeV. }
\label{figs}       
\end{figure}

Setting $N=3$, our holographic estimates for $\frac{s}{T^3}$ are summarized in Fig.~\ref{figs}, where each curve starts at the corresponding critical deconfinement temperature. Qualitatively, these curves compare reasonably well with \cite[Fig.~10]{Bali:2014kia}, at least for moderate values of the magnetic field. For larger magnetic fields, we notice that $\frac{s}{T^3}$ initially overshoots the (magnetic field independent) asymptotical Stefan-Boltzmann value, in contrast with corresponding lattice data.

\section{Outlook}
In this short Letter, we have shown how to generalize the potential reconstruction method to the case of multiple quark flavours in the quenched case. We have included a background magnetic field via slightly gauging the $U(1)\times U(1)$ invariance of the (unbroken) underlying $U(2)$ vector symmetry.

Next to the confirmed inverse magnetic catalysis for what concerns the deconfinement temperature and a reasonable description of the entropy, another test of the novel model would be to study the vector meson spectrum, even at vanishing magnetic field, since the coupling function is, in contrast with earlier work, now internally determined rather than freely chosen.

In order to further improve the model, we should take into account the backreaction of the added flavours on the gravitational background, i.e.~to unquench the setup. We should also include explicitly the (broken) chiral symmetry via appropriate gauge fields, thereby fully incorporating the global flavour symmetries of QCD. Once this is properly done, the complete thermodynamics, chiral dynamics, including the chiral transition, etc.~can be scrutinized. The interplay with holographic entanglement entropy, see \cite{Dudal:2016joz,Arefeva:2020uec}, would also be an interesting further application.

We hope to come back to all of these issues in the near future.

\begin{acknowledgements}
We thank the Guest Editors for their kind invitation to contribute to this Topical Issue.

\end{acknowledgements}

\end{document}